\documentclass[aps, prd, twocolumn, lengthcheck, superscriptaddress, showpacs, letterpaper, nofootinbib]{revtex4-1}

\usepackage{amsmath,amssymb}
\usepackage{graphicx,bm}
\usepackage{slashed}
\usepackage{dcolumn}
\usepackage{epstopdf}
\usepackage{ulem} %% for strike-through
\usepackage[usenames]{color}
\usepackage{mathrsfs}
\usepackage{float}
\usepackage{diagbox}
\usepackage{multirow}

\allowdisplaybreaks

\begin{document}

\title{Pentaquark states from bound state approach with chiral partner structure}

\author{Ben-Teng Zhang}

\affiliation{Center for Theoretical Physics and College of Physics, Jilin University, Changchun,
130012, China}

\author{Jun-Shuai Wang}

\affiliation{Center for Theoretical Physics and College of Physics, Jilin University, Changchun,
130012, China}

\author{Yong-Liang Ma}
\email{yongliangma@jlu.edu.cn}
\affiliation{Center for Theoretical Physics and College of Physics, Jilin University, Changchun,
130012, China}

\date{\today}
%%%%%%%%%%%%%%%%%%%%%%%%%%%%%%%
\begin{abstract}
The observation of the two states $P_c(4380)$ and $P_c(4445)$ [Phys.Rev.Lett. 115 (2015) 072001] solids the existence of the pentaquark state in Nature. In this work, we study the spectrum of the pentaquark states including one or two heavy quarks based on the bound state approach by binding the heavy-light mesons to the nucleon as a soliton in an effective Lagrangian approach. By regarding the $H$ doublet and $G$ doublet as chiral partners to each other and coupling them to light mesons with a minimal derivative, the chiral partner structure of the pentaquark states is calculated. We find that the bound state approach prefers to arrange $P_c^+(4440)$ and $P_c(4457)$ into a heavy quark doublet $(\frac{1}{2}^-,\frac{3}{2}^-)$ so both decay to $J/\psi p$ through the $S$-wave. The predicted spectrum shown in this work serves as a guide for the experimental hunting of more pentaquark states.

\end{abstract}

%\pacs{
%12.39.Dc   % Skyrmions
%12.39.Fe   % Chiral Lagrangians
%21.65.Ef   % Symmetry energy
%}

\maketitle

%%%%%%%%%%%%%%%%%%%%%%%%%%%%%%%%%%%%%%%%%%%%%%%%%%%%%%%%%%%%%%%%%%%%%%%

\section{introduction}
%\sect{\bf Introduction}
\label{sec:intro}

In 2015, two peaks which were named $P_c^+(4380)$ and $P_c^+(4450)$ were observed at LHCb collaboration in the invariant mass distribution of $J/\psi p$ in the decay $\Lambda_b^0 \to J/\psi K^- p$ ~\cite{Aaij:2015tga}. Very recently, after a closer analysis of the data, the LHCb collaboration reported a new enhancement, the $P_c^+(4312)$ was found and the previously reported state $P_c^+(4450)$ split into two structures, $P_c^+(4440)$ and $P_c^+(4457)$~\cite{Aaij:2019vzc}. With respect to their decay products, one can easily conclude that these newly observed $P_c^+$ states consist of at least five quarks, i.e., they are hidden charm pentaquark states. So far, we don't definitely know any other quantum number of these pentaquark states except isospin which can be fixed from their decay products $J/\Psi p$ to be $I=1/2$.

Since the report of the two states, $P_c(4380)$ and $P_c(4445)$~\cite{Aaij:2015tga}, a lot of theoretical studies have been performed to explain their characters. Some of these suggest that these pentaquarks may be interpreted as molecular states~\cite{Chen:2015loa,Chen:2015moa,Karliner:2015ina,Roca:2015dva,He:2015cea,Meissner:2015mza,Shen:2016tzq,Liu:2019tjn}, tightly bound pentaquark states~\cite{Maiani:2015vwa,Wang:2015ava,Lebed:2015tna,Wang:2015epa}, just the kinematical effects of the rescattering from $\chi_{c1} p \to J/\psi p$~\cite{Guo:2015umn} or threshold effect~\cite{Liu:2015fea}. There's another idea to account for the masses and quantum numbers of the pentaquarks using the bound state approach in which the pentaquark states are composite states formed by a baryon and two heavy-light mesons~\cite{Scoccola:2015nia}. More references can be found in, e.g., reviews~\cite{Chen:2016qju,Ali:2017jda,Guo:2017jvc,Liu:2019zoy}.

In this work, we study the spectrum of pentaquark states by using the bound state approach by binding two heavy-light mesons with chiral partner structure to the nucleon as the soliton $S$.  In particular, we regard the $H$ doublet and $G$ doublet ($H$ and $G$ doublets will be specified later) as chiral partners~\cite{Nowak:1992um,Bardeen:1993ae}, so the composite open charm pentaquark states have two scenarios $\bar{H}S$ and $\bar{G}S$~\cite{Harada:2012dm} while the composite hidden charm pentaquark states have four scenarios, $HS\bar{H}$, $HS\bar{G}$, $GS\bar{H}$, $GS\bar{G}$.

In the literature, the heavy pentaquark states have been studied by using the bound approach ~\cite{Oh:1994yv,Oh:1994np,Harada:1997bk,Wu:2004wg,Schechter:1995vr,Scoccola:2015nia} including the chiral partner structure~\cite{Nowak:2004jg,Harada:2012dm}. Here, we extend what we did in Ref.~\cite{Harada:2012dm} to study the hidden heavy quarkonium pentaquark states after a first revisit the open charm pentaquark states.

What we found here is that, the states $P_c^+(4440)$ and $P_c^+(4457)$ in the updated data~\cite{Aaij:2019vzc} can be arranged in the heavy quark doublet $(\frac{1}{2}^-,\frac{3}{2}^-)$ and therefore both decay to $J/\psi p$ through the $S$-wave. Due to the chiral partner structure of the constituent heavy-light mesons, we predict more pentaquark states which can decay into open charmed mesons. From the present results one can easily obtain the pentaquark states including bottom quark.

This paper is organized as follows: We illustrate our theoretical framework and results of the open charm pentaquark states in Sec.~\ref{sec:Fram}. Our prediction of the hidden charm pentaquark states are calculated in Sec.~\ref{sec:Hidden}. We give our discussion of the present calculation in Sec.~\ref{sec:Dis}.

\section{Theoretical framework and open charm pentaquark states}

\label{sec:Fram}

We start from Ref.~\cite{Harada:2012dm} which investigated the chiral partner structure of the pentaquark states including one heavy quark based on the bound state approach by binding the heavy-mesons in the $H$ doublet or $G$ doublet to the nucleon as the soliton $S$ in an effective Lagrangian for the pseudoscalar and vector mesons based on hidden local symmetry~\cite{Harada:2003jx}. We will only cover the part relevant to the present work.

We introduce the charmed heavy-light meson doublets $H$ and $G$ with quantum numbers $J^P=(0^-, 1^-)$ and $J^P = (0^+, 1^+)$, respectively, into the model. In terms of the physical states and the notation of PDG~\cite{Tanabashi:2018oca}, they are expressed as
\begin{eqnarray}
H & = & \frac{1 + v\hspace{-0.15cm}\slash}{2}\left(D^{\ast;\mu}\gamma_\mu + i D \gamma_5\right),\nonumber\\
G & = & \frac{1 + v\hspace{-0.15cm}\slash}{2}\left(D^{\prime;\mu}_1\gamma_\mu\gamma_5 + D^\ast_0\right),
\label{eq:HeavyMeson}
\end{eqnarray}
which have charm number $c={}+1$. Regarding the $H$ doublet and $G$ doublet as chiral partners to each other~\cite{Nowak:1992um,Bardeen:1993ae}, we write the effective Lagrangian describing the interaction between the heavy-light mesons and the light mesons as~\cite{Nowak:1992um,Bardeen:1993ae,Harada:2012dm}~\footnote{The coupling between the $H$ doublet and $G$ doublet is
\begin{eqnarray}
{\cal L}_{\rm GH} = g_{GH} {\rm Tr}\left[H \gamma^\mu\gamma_5 \hat{\alpha}_{\perp\mu} \bar{G}\right] + {\rm h.c.}.
\end{eqnarray}
Therefore, in the heavy quark limit, its contribution to the bound state vanishes in the rest frame of the soliton~\cite{Nowak:2004jg}.
}
\begin{eqnarray}
{\cal L}_{\rm heavy} & = & {\rm Tr}\left[G(iv\cdot\tilde{D})\bar{G}\right] - {\rm Tr}\left[H(iv\cdot\tilde{D})\bar{H}\right]\nonumber\\
& & {} + g_A {\rm Tr}\left[H\gamma^{\mu}\gamma_5\hat{\alpha}_{\perp\mu}\bar{H}\right]-g_A{\rm Tr}\left[G\gamma^{\mu}\gamma_5\hat{\alpha}_{\perp\mu}\bar{G}\right],\nonumber\\
\label{eq: HHChPT}
\end{eqnarray}
where
\begin{eqnarray}
\hat{\alpha}_{\parallel\mu} & = & \frac{1}{2i}(D_{\mu}\xi_R\cdot\xi_R^\dagger+D_{\mu}\xi_L\cdot\xi_L^\dagger),\nonumber\\
\hat{\alpha}_{\perp\mu} & = & \frac{1}{2i}(D_{\mu}\xi_R\cdot\xi_R^\dagger-D_{\mu}\xi_L\cdot\xi_L^\dagger),
\end{eqnarray}
and the covariant derivative $\tilde{D}_\mu$ is defined as
\begin{eqnarray}
D_\mu % & = & \partial_{\mu}-iV_{\mu}\nonumber\\
& = & \partial_{\mu}-i\frac{g}{2}(\bm{\tau}\cdot \bm{\rho}_{\mu}+\omega_{\mu}),\nonumber\\
\tilde{D}_\mu % & = & \partial_{\mu}-iV_{\mu}-i\kappa\hat{\alpha}_{\parallel\mu}\nonumber\\
& = & D_\mu - i\kappa\hat{\alpha}_{\parallel\mu}.
\end{eqnarray}
In Lagrangian~\eqref{eq: HHChPT}, $\xi^{\dagger}_L = \xi_R = \sqrt{U}=e^{i\pi/f_{\pi}}$, the equality of the coupling constants in the third and fourth terms are due to the chiral partner structure, $f_\pi$, $g$, $g_A$ and $\kappa$ are parameters fixed in Ref.~\cite{Harada:2012dm}.

In the bound state approach, due to the spin-isospin correlation in the adopted ``hedgehog" ansatz for soliton configuration~\cite{Skyrme:1961vq}, and the conservation of the heavy quark spin, the equation of motion of heavy-light meson field is invariant the rotation of the `` light grand spin"
\begin{eqnarray}
\bm{g}&=\bm{L}+\bm{J}_{\rm light}+\bm{I}_{\rm light} = \bm{L}+\bm{K},
\label{eq:lgrand}
\end{eqnarray}
where $\bm{L}$ is the ordinary orbital angular momentum operator between the soliton and heavy-light meson, $\bm{I}_{\rm light}$ is the isospin operator of the heavy-light meson field and, $\bm{J}_{\rm light}$ is the spin operator of the light degree of freedom of the heavy-light meson which has the eigenvalue $1/2$ for both $H$ and $G$ doublets. In the bound state approach, for the quantum number $l$ of operator $\bm{L}$, the $HS (\bar{H}S)$ bound states have parity $\pi_H = (-1)^l$ while the $GS (\bar{G}S)$ bound states have parity $\pi_G={}-(-1)^l$. In this work we will consider both the large $N_c$ limit and heavy quark limit, i.e., both soliton and heavy-light mesons have infinite mass. Therefore, we will not consider the orbital excitation between the constituents, i.e., take $l = 0$.

From Lagrangian \eqref{eq: HHChPT}, one can express the potential $V_H$ between $H$ doublet and soliton and potential $V_G$ between $G$ doublet and soliton as
\begin{eqnarray}
V_H & = & \frac{1}{2}(1 + \kappa)g\omega(0)+g_AF^{'}(0)\left[k(k+1)-\frac{3}{2}\right],\nonumber\\
V_G & = & \frac{1}{2}(1 + \kappa)g\omega(0)-g_AF^{'}(0)\left[k(k+1)-\frac{3}{2}\right],
\label{eq:BE}
\end{eqnarray}
where $k$ is the eigenvalue of the operator $\bm{K} = \bm{J}_{\rm light}+\bm{I}_{\rm light}$ defined in Eq.~\eqref{eq:lgrand}. In Eq.~\eqref{eq:BE}, $F(0)$ and $\omega(0)$ are, respectively, the static profile functions of pion and omega at origin. Their values depend on the profiles of the soliton configuration at origin, therefore the model of soliton. Here, to consistent with the present work, we take the values from Ref.~\cite{Ma:2012zm}

The state with definite angular momentum and isospin can be generated by collective quantization. The collective rotation gives an additional contribution to the Largrangian~\cite{Callan:1985hy}
\begin{equation}
\delta{\cal L}_{\rm coll}=\frac{1}{2}\mathcal{I}\Omega^2+{\bm{\Theta}}\cdot\bm{\Omega},
\end{equation}
where $\mathcal{I}$ is the moment of inertia of the soliton and $\bm{\Omega}$ is the angular velocity of the collective rotation. $\bm{\Theta}$ is the isospin of the heavy-light mesons interacting with the nucleon as soliton~\cite{Oh:1994yv}. The angular momentum computed from this collective rotation is
\begin{eqnarray}
\bm{J}^{\rm rot} =\dfrac{\partial\delta{\cal L}_{\rm coll}}{\partial\bm{\Omega}}=\mathcal{I}\bm{\Omega}+{\bm{\Theta}},
\end{eqnarray}
where $\bm{J}^{\rm rot}$ and the total isospin $\bm{I}$ satisfying $I=J^{\rm rot}$~\cite{Oh:1994yv}. Now, the total spin of the light degrees of freedom in the heavy baryon (bound state) is expressed as
\begin{equation}
\bm{j}=\bm{J}^{\rm rot} + \bm{K}.
\end{equation}
With respect to the heavy quark spin, the spin operator of the composite heavy baryon is expressed as $\bm{J}_B=\bm{j}+\bm{S}_Q$ with eigenvalues $j_B=j\pm1/2$.

By the standard Legendre transformation, one can get the rotation energy $H_{\rm coll}$ as
\begin{equation}
H_{\rm coll}=\frac{1}{2\mathcal{I}}\left[\left(1-\chi(k)\right)\bm{I}^2+\chi(k)\bm{j}^2-\chi(k)\bm{K}^2+\frac{3}{4}\right],
\end{equation}
where
\begin{eqnarray}
\chi(k) & = &
\left\{
  \begin{array}{ll}
    \frac{[k(k+1)+3/4-j_{\rm light}(j_{\rm light}+1)]}{2k(k+1)}, & \hbox{$k\neq0$;} \\
    0, & \hbox{ $k=0$.}
  \end{array}
\right.
\label{eq:chik}
\end{eqnarray}

With the above discussion, we then finally obtain the bound state mass as
\begin{eqnarray}
M_{\rm BS}=M_{\rm sol}+\bar{M}_{H,G}+V_{H,G}+H_{\rm coll}.
\label{eq:BSmass1}
\end{eqnarray}
In this expression, $\bar{M}_{H,G}$ are the spin-averaged heavy-light meson masses with
\begin{eqnarray}
\bar{M}_H & = & \frac{m_{D_0}+3m_{D^*}}{4}= 1.98~{\rm GeV}, \nonumber\\
\bar{M}_G & = & \frac{m_{D_0^*}+3m_{D_1}}{4}= 2.41~{\rm GeV}.
\end{eqnarray}
Using the empirical values of the baryon masses $m_N=0.94~{\rm GeV}, m_{\Delta}=1.23~{\rm GeV}$, we obtain the following values of the parameters
\begin{eqnarray}
& & M_{\rm sol}=0.868~{\rm GeV},\;\; 1/\mathcal{I}=0.193~{\rm GeV}, \;\; g =4.74.
\end{eqnarray}
By using the $D^\ast \to D \pi$, we obtain $|g_A|\simeq0.56$~\cite{Harada:2012km}. In combination with $m_{\Lambda_c}=2.29~{\rm GeV}$ which can be regarded as a $HS$ bound state in the bound state approach, we can fix $\kappa={}-0.83$ with $g_A=0.56$~\cite{Harada:2012dm}. With this parameter choice and the profile functions of the light mesons calculated from the HLS upto the next to leading order~\cite{Ma:2012zm}, the binding energy between the heavy-light mesons in the $H$ doublet and $G$ doublet and soliton can be obtained as
\begin{eqnarray}
V_H & = &{} -0.03 + 0.35\left[k(k+1)-\frac{3}{2}\right]~{\rm GeV},\nonumber\\
V_G & = &{} -0.03 - 0.35\left[k(k+1)-\frac{3}{2}\right]~{\rm GeV}.
\end{eqnarray}
Substituting $v_{\mu}$ with ${} -v_{\mu}$, repeating the previous calculation, one can obtain the binding energy between the heavy-light mesons in the anti-$H$ doublet $(c={}-1)$ and anti-$G$ doublet $(c={}-1)$ and soliton as
\begin{eqnarray}
V_{\bar{H}} & = & 0.03 - 0.35\left[k(k+1)-\frac{3}{2}\right]~{\rm GeV},\nonumber\\
V_{\bar{G}} & = & 0.03 + 0.35\left[k(k+1)-\frac{3}{2}\right]~{\rm GeV}.
\end{eqnarray}
Such kind of bound states made of a soliton and the anti-$H$ doublet or anti-$G$ doublet are pentaquark states with one anti-charm quark. We list in
Table~\ref{table:BE} the binding energy for  quantum numbers $k =0,1$.

%\begin{widetext}
%\begin{sidewaystable}[h]
\begin{table}[!htp]
\centering
%\fontsize{9.5}{9}\selectfont
\caption{Binding energy between the heavy-light meson and soliton (in unit GeV).}
\label{table:BE}
\label{Table}
\begin{tabular}{c|cc}
\hline
\hline
$~$ & ~~~ $k$ ~~ & Binding energy
\cr
\hline
\multirow{2}{*}{$H$ doublet}
& $0$ & ${}-0.556$ \cr
& $1$ & $0.145$
\cr
\hline
\multirow{2}{*}{$\bar{H}$ doublet}
& $0$ & $0.556$
\cr
& $1$ & ${}-0.145$
\cr
\hline
\multirow{2}{*}{$G$ doublet}
& $0$ & $0.496$\cr
& $1$ & ${}-0.205$
\cr
\hline
\multirow{2}{*}{$\bar{G}$ doublet}
& $0$ & ${}-0.496$
\cr
& $1$ & $0.205$
\cr

\hline
\hline
\end{tabular}
%\end{sidewaystable}
\end{table}
%\end{widetext}

The results shown in Table~\ref{table:BE} tell us that, for the $H$ doublet with $c={}+1$, the $k=0$ channel forms the first bound state while for the $G$ doublet with $c={}+1$, the $k=1$ channel forms the first bound state. However, for the bound states with anti-charm quark, i.e., the $c={}-1$ sector,  the $\bar{H}$ doublet forms the first bound state in the $k={}1$ channel while the $\bar{G}$ doublet forms the first bound state in the $k={}0$ channel. These states are regarded as pentaquark states with $c={}-1$.

\begin{table}[!htp]
\centering
%\fontsize{9.5}{9}\selectfont
\caption{Lowest-lying pentaquark states with one anticharm quark (in unit GeV).}
\label{table:mass1c}
\label{Table}
\begin{tabular}{cccccc}
\hline
\hline
Bound state & ~ $I$ &~ $j$ & $I(j_B^P)$ ~ & Candidates & mass \cr
\hline
$\bar{H}S$ & ~ $0$ &~ $1$ & ~$(\frac{1}{2}^+,\frac{3}{2}^+)$ ~ & $\Theta_c(\frac{1}{2}^+),\Theta_c(\frac{3}{2}^+)$ &~ ${}2.745$
\cr
$\bar{G}S$ & ~ $0$ &~ $0$ &~ $\frac{1}{2}^-$ ~ & $\Theta_c(\frac{1}{2}^-)$ &~ ${}2.778$
\cr
\hline
\hline
\end{tabular}
%\end{sidewaystable}
\end{table}

We list the predicted spectrum of the pentaquark states in Table \ref{table:mass1c}. From this table, we see that, although the mass difference between the $H$ doublet and $G$ double is about $\sim 430$~MeV, the bound states $\bar{H}S$ and $\bar{G}S$ have the close mass. This is because, the $\bar{G}$ doublet is more deeply bound to the soliton. In stark contrast to the standard scenario of chiral partner structure in hadron physics where the mass splitting between chiral partners is $\sim 430$~MeV for nonstrange hadrons~\cite{Nowak:1992um,Bardeen:1993ae}, the mass splitting between the chiral partners of the pentaquark predicted from the bound state approach is about $\sim 30$~MeV. Another point different from the standard chiral partner scenario is that, the $\bar{H}S$ bound states form a heavy quark doublet with quantum numbers $(\frac{1}{2}^+,\frac{3}{2}^+)$ but the corresponding chiral partner is a heavy quark singlet with quantum numbers $\frac{1}{2}^-$. It should be noticed that, due to the deep binding energy in the $\bar{G}S$ channel, all the states are below the threshold of $\bar{D}p$, therefore, these pentaquark states decay through weak force so that are narrow width states, not easy to observe.

\section{Hidden charm pentaquark states}

\label{sec:Hidden}

After the discussion of the pentaquark states with one anti-charm quark, we extend the bound state approach to discuss the hidden charm pentaquark states. In this case, different from the pentaquark states with one anti-charm quark we discussed in above,  two heavy-light mesons --- one charm and the other anticharm --- should be bound to the soliton. In such a case, the rotational energy reads~\cite{Rho:1990uy}
\begin{equation}
\delta {\cal L}_{\rm coll}=\frac{1}{2}\mathcal{I} \Omega^2 + {\bm \Theta}_1\cdot{\bm \Omega} + {\bm \Theta}_2\cdot {\bm \Omega}
\end{equation}
with $\bm{\Theta}_i$ being the isospin of the $i$-th heavy-light meson. Therefore, the total spin of the light degrees of freedom in the bound state --- including two heavy-light mesons ---  is defined as
\begin{equation}
\label{31}
\bm{j}_l=\bm{J}^{\rm rot}+\bm{K}_1+\bm{K}_2,
\end{equation}
with $\bm{K}_i$ being the sum of the light-spin and isospin of the $i$-th heavy-light mesons. By including the heavy quark spin, the spin operator for the bound pentaquark state $P_c$ is expressed as  $\bm{j}_{P_c}=\bm{j}_l+\bm{S}_{Q_1}+\bm{S}_{Q_2}$. Therefore, in the bound state approach, different from the pentaquark states including one heavy quark, the pentaquark states with two heavy quarks are arranged into two heavy quark multiplets. In the present hidden charm pentaquark case, the bound states made of the $i$-th and the $j$-th heavy-light mesons have parity $\pi_i\pi_j$ .

By defining the total $K$-spin of the light degrees of freedom in the system $\bm{K}_l=\bm{K}_1+\bm{K}_2$, the $H_{\rm coll}$ can be expressed as~\cite{Rho:1990uy}
\begin{eqnarray}
H_{\rm coll} & = & \dfrac{1}{2\mathcal{I}}\Bigg\{I(I+1)+3/2 \nonumber\\
& &\qquad {} +\chi_1\chi_2[k_l(k_l+1) -k_1(k_1+1)-k_2(k_2+1)]\nonumber\\
& &\qquad{} +\left[\frac{\chi_+}{2}+\frac{\chi_-}{2}\dfrac{k_1(k_1+1)-k_2(k_2+1)}{k_l(k_l+1)}\right]\nonumber\\
& &\quad\qquad{} \times[j_l(j_l+1)-I(I+1)-k_l(k_l+1)] \Bigg\},\nonumber\\
\end{eqnarray}
where $\chi_{\pm} = \chi_1 \pm \chi_2$ with $\chi_i$ given by Eq.~\eqref{eq:chik} for the $i$-th heavy-light meson. Then we finally obtain the mass of the bound state --- pentaquark state --- as
\begin{equation}
M=M_{\rm sol} + \bar{M}_1 + \bar{M}_2 + V_1 + V_2 + H_{\rm coll},
\end{equation}
where $\bar{M}_i$ is the spin-averaged mass of the $i$-th heavy-light meson in the bound state, $V_i$ is the binding energy between the soliton and the $i$-th heavy-light meson which is given in Table~\ref{table:BE}.

From Table~\ref{table:BE}, we see that, for the $c={}+1$ meson, the $H$ doublet bound to soilton in the $k^{\pi_H}=0^-$ channel while the $G$ doublet bound to soliton in the $k^{\pi_G}=1^+$ channel. However, the $c={}-1$ meson, the $\bar{H}$ doublet bound to soliton in the  $k^{\pi_{\bar{H}}}=1^-$ channel while the $\bar{G}$-doublet bound to soliton in the $k^{\pi_{\bar{G}}}=0^+$ channel.  The parity of the bound state pentaquark state is $\pi=\pi_i\pi_{j}$ with $\pi_i$ being the parity of the $i$-th heavy-light meson in the bound state. With respect to this fact, we list our result in Table~\ref{table:HC5q} together with the corresponding masses and quantum numbers.

%\begin{widetext}
%\begin{sidewaystable}[h]
\begin{table}[!htp]
\centering
%\fontsize{9.5}{9}\selectfont
\caption{Spectrum of the hidden-charm pentaquark states $M_1 S \bar{M}_2$ (in unit MeV).}
\label{table:HC5q}
\label{Table}
\begin{tabular}{c|cccccccc}
\hline
\hline
 ~~ & ~ $I$ & ~ $k_1^{\pi_1} $ & $k_2^{\pi_2} ~ $ & ~ $k^\pi $ ~ & $j_l^\pi$ ~~~ & $j_{P_c}^P$ ~& Mass
\cr
\hline
\multirow{2}{*}{$HS\bar{H}$}
& ~ $\frac{1}{2}$ & $0^+$ & $1^+$ & $1^+$ & $\frac{1}{2}^+$ & $\frac{1}{2}^+, (\frac{1}{2}^+,\frac{3}{2}^+)$ & $4.140$\cr
& $~$ & $~$ & $~$ & $~$ & $\frac{3}{2}^+$ & $\frac{3}{2}^+, (\frac{1}{2}^+,\frac{3}{2}^+,\frac{5}{2}^+)$ & $4.238$
\cr
\hline
\multirow{1}{*}{$HS\bar{G}$}
& ~ $\frac{1}{2}$ & $0^+$ & $0^-$ & $0^-$ & $\frac{1}{2}^-$ & $\frac{1}{2}^-, (\frac{1}{2}^-,\frac{3}{2}^-)$ & $4.417$\cr
%& $~$ & $~$ & $~$ & $~$ & $\frac{1}{2}^+, \frac{3}{2}^+$ & $4.238$
%\cr
\hline
\multirow{5}{*}{$GS\bar{H}$}
& ~ $\frac{1}{2}$ & $1^-$ & $1^+$ & $2^-$ & $\frac{5}{2}^-$ & $\frac{5}{2}^-, (\frac{3}{2}^-,\frac{5}{2}^-,\frac{7}{2}^-)$ & $5.264$\cr
& $~$ & $~$ & $~$ & $~$ & $\frac{3}{2}^-$ & $\frac{3}{2}^-, (\frac{1}{2}^-,\frac{3}{2}^-,\frac{5}{2}^-)$ & $5.021$
\cr
& ~ $~$ & $~$ & $~$ & $1^-$ & $\frac{3}{2}^-$ & $\frac{3}{2}^-, (\frac{1}{2}^-,\frac{3}{2}^-,\frac{5}{2}^-)$ & $5.118$\cr
& $~$ & $~$ & $~$ & $~$ & $\frac{1}{2}^-$ & $\frac{1}{2}^-, (\frac{1}{2}^-,\frac{3}{2}^-)$ & $4.972$
\cr
& ~ $~$ & $~$ & $~$ & $0^-$ & $\frac{1}{2}^-$ & $\frac{1}{2}^-, (\frac{1}{2}^-,\frac{3}{2}^-)$ & $5.021$\cr
\hline
\multirow{2}{*}{$GS\bar{G}$}
& ~ $\frac{1}{2}$ & $1^-$ & $0^-$ & $1^+$ & $\frac{3}{2}^+$ & $\frac{3}{2}^+, (\frac{1}{2}^+,\frac{3}{2}^+,\frac{5}{2}^+)$ & $5.297$\cr
& $~$ & $~$ & $~$ & $~$ & $\frac{1}{2}^+$ & $\frac{1}{2}^+, (\frac{1}{2}^+,\frac{3}{2}^+)$ & $5.005$
\cr

\hline
\hline
\end{tabular}
%\end{sidewaystable}
\end{table}
%\end{widetext}

From Table~\ref{table:HC5q}, we find that all the predicted states are above the threshold of the $J/\Psi p$, i.e., $M_{J/\Psi p} \simeq 4.04$~GeV. If the decay products $J/\Psi p$ of $P_c$ are determined in $S$-wave, it can be identified with $HS\bar{G}$ bound states. However, if the products $J/\Psi p$ are in the $P$-wave, it should be the $HS\bar{H}$ bound states.

Comparing our prediction with the observation, we found that it is reasonable to identify the  $P_c^+(4440)$ and $P_c^+(4457)$ with the $HS\bar{G}$ bound state and arrange them in the heavy quark doublet $(\frac{1}{2}^-, \frac{3}{2}^-)$. The mass splitting between the $P_c^+(4440)$ and $P_c^+(4457)$, $\sim 20$~MeV, is due to the heavy quark symmetry breaking effect, like that of the heavy-light mesons in the $G$ doublet. In the present calculation, it is difficult for us to identify the quantum numbers of $P_c(4312)$ by simply comparing the mass.

We next discuss the chiral partner structure in the present picture. In the present bound state approach to the pentaquark states, it is more difficult to identify the chiral partner than that in the heavy-light meson sector and the pentaquark state including one heavy quark. A possible scheme, probably more reasonable one, is to clarify the bound states $HS\bar{H}$ and $HS\bar{G}$ as chiral partners to each other and, $GS\bar{H}$ and $GS\bar{G}$ as another chiral partners. In this scenario, the chiral partner structure arises from anti-heavy-light meson constituent in the bound state. In addition, due to contribution to the rotation energy from the combination of the quantum numbers $k$ and $j_l$, the mass splitting between the chiral partners is subtle.

\section{Summary and discussion}

\label{sec:Dis}

In this work, we studied the spectrum of heavy quark pentaquark states from the bound state approach with chiral partner structure. We take both the large $N_c$ limit and the heavy quark limit, therefore the constituents can be regarded as infinitely heavy objects. With respect to this, we will not consider the radial excitation of the constituents.

From our calculation, it seems reasonable to identify the $P_c^+(4440)$ and $P_c^+(4457)$ as heavy quark doublet with quantum numbers $(\frac{1}{2}^-,\frac{3}{2}^-)$. Their chiral partners have complicated structure and have about $200~$MeV smaller masses. These are the main conclusions in the this work. In addition, due to the combinations of the heavy-light mesons in the bound states, many other states are predicted. These states may give a guide for the future hinting of more pentaquark states although, after more precise calculations, for example, including the finite heavy quark mass corrections, the masses predicted here may be slightly changed. Different from what observed so far, these states are above threshold of $D\bar{D}N$ with $N$ being the nucleon so that can be observed in the channels including open charm mesons.

From the present numerical results, one can easily estimate the pentaquark states including bottom quarks, i.e., the states have valence quarks $\bar{b}qqqq, \bar{b}cqqq, \bar{c}bqqq$ and $\bar{b}bqqq$ with $q$ being the light quark. This can be straight forwardly done by substituting the spin-averaged mass of the charmed heavy-light meson doublet with the corresponding one of the bottom heavy-light meson doublet. Since the potential and collective rotation energy calculated in the bound state approach only concern the light degrees freedom, what one should do is to add the mass difference between the bottom doublet and the corresponding charmed doublet. We will not list the numbers here since it is a trivial procedure.

%\footnote{\cmh{Strange pentaquark?}}

\subsection*{Acknowlegments}

The work of Y.~L. Ma was supported in part by National Science Foundation of China (NSFC) under Grant No. 11875147 and 11475071.

%%%%%%%%%%%%%%%%%%%%%%%%%%%%%%%%%%%%%%%%%%%%%%%%%%%%%%%%%%%%%%%%%%%%%%%%%
%\appendix

\bibliography{RefChiralPenta}

\begin{thebibliography}{37}%
\makeatletter
\providecommand \@ifxundefined [1]{%
 \@ifx{#1\undefined}
}%
\providecommand \@ifnum [1]{%
 \ifnum #1\expandafter \@firstoftwo
 \else \expandafter \@secondoftwo
 \fi
}%
\providecommand \@ifx [1]{%
 \ifx #1\expandafter \@firstoftwo
 \else \expandafter \@secondoftwo
 \fi
}%
\providecommand \natexlab [1]{#1}%
\providecommand \enquote  [1]{``#1''}%
\providecommand \bibnamefont  [1]{#1}%
\providecommand \bibfnamefont [1]{#1}%
\providecommand \citenamefont [1]{#1}%
\providecommand \href@noop [0]{\@secondoftwo}%
\providecommand \href [0]{\begingroup \@sanitize@url \@href}%
\providecommand \@href[1]{\@@startlink{#1}\@@href}%
\providecommand \@@href[1]{\endgroup#1\@@endlink}%
\providecommand \@sanitize@url [0]{\catcode `\\12\catcode `\$12\catcode
  `\&12\catcode `\#12\catcode `\^12\catcode `\_12\catcode `\%12\relax}%
\providecommand \@@startlink[1]{}%
\providecommand \@@endlink[0]{}%
\providecommand \url  [0]{\begingroup\@sanitize@url \@url }%
\providecommand \@url [1]{\endgroup\@href {#1}{\urlprefix }}%
\providecommand \urlprefix  [0]{URL }%
\providecommand \Eprint [0]{\href }%
\providecommand \doibase [0]{http://dx.doi.org/}%
\providecommand \selectlanguage [0]{\@gobble}%
\providecommand \bibinfo  [0]{\@secondoftwo}%
\providecommand \bibfield  [0]{\@secondoftwo}%
\providecommand \translation [1]{[#1]}%
\providecommand \BibitemOpen [0]{}%
\providecommand \bibitemStop [0]{}%
\providecommand \bibitemNoStop [0]{.\EOS\space}%
\providecommand \EOS [0]{\spacefactor3000\relax}%
\providecommand \BibitemShut  [1]{\csname bibitem#1\endcsname}%
\let\auto@bib@innerbib\@empty
%</preamble>
\bibitem [{\citenamefont {Aaij}\ \emph {et~al.}(2015)\citenamefont {Aaij} \emph
  {et~al.}}]{Aaij:2015tga}%
  \BibitemOpen
  \bibfield  {author} {\bibinfo {author} {\bibfnamefont {R.}~\bibnamefont
  {Aaij}} \emph {et~al.} (\bibinfo {collaboration} {LHCb}),\ }\href {\doibase
  10.1103/PhysRevLett.115.072001} {\bibfield  {journal} {\bibinfo  {journal}
  {Phys. Rev. Lett.}\ }\textbf {\bibinfo {volume} {115}},\ \bibinfo {pages}
  {072001} (\bibinfo {year} {2015})},\ \Eprint
  {http://arxiv.org/abs/1507.03414} {arXiv:1507.03414 [hep-ex]} \BibitemShut
  {NoStop}%
%%CITATION = ARXIV:1507.03414;%%
\bibitem [{\citenamefont {Aaij}\ \emph {et~al.}(2019)\citenamefont {Aaij} \emph
  {et~al.}}]{Aaij:2019vzc}%
  \BibitemOpen
  \bibfield  {author} {\bibinfo {author} {\bibfnamefont {R.}~\bibnamefont
  {Aaij}} \emph {et~al.} (\bibinfo {collaboration} {LHCb}),\ }\href {\doibase
  10.1103/PhysRevLett.122.222001} {\bibfield  {journal} {\bibinfo  {journal}
  {Phys. Rev. Lett.}\ }\textbf {\bibinfo {volume} {122}},\ \bibinfo {pages}
  {222001} (\bibinfo {year} {2019})},\ \Eprint
  {http://arxiv.org/abs/1904.03947} {arXiv:1904.03947 [hep-ex]} \BibitemShut
  {NoStop}%
%%CITATION = ARXIV:1904.03947;%%
\bibitem [{\citenamefont {Chen}\ \emph
  {et~al.}(2015{\natexlab{a}})\citenamefont {Chen}, \citenamefont {Liu},
  \citenamefont {Li},\ and\ \citenamefont {Zhu}}]{Chen:2015loa}%
  \BibitemOpen
  \bibfield  {author} {\bibinfo {author} {\bibfnamefont {R.}~\bibnamefont
  {Chen}}, \bibinfo {author} {\bibfnamefont {X.}~\bibnamefont {Liu}}, \bibinfo
  {author} {\bibfnamefont {X.-Q.}\ \bibnamefont {Li}}, \ and\ \bibinfo {author}
  {\bibfnamefont {S.-L.}\ \bibnamefont {Zhu}},\ }\href {\doibase
  10.1103/PhysRevLett.115.132002} {\bibfield  {journal} {\bibinfo  {journal}
  {Phys. Rev. Lett.}\ }\textbf {\bibinfo {volume} {115}},\ \bibinfo {pages}
  {132002} (\bibinfo {year} {2015}{\natexlab{a}})},\ \Eprint
  {http://arxiv.org/abs/1507.03704} {arXiv:1507.03704 [hep-ph]} \BibitemShut
  {NoStop}%
%%CITATION = ARXIV:1507.03704;%%
\bibitem [{\citenamefont {Chen}\ \emph
  {et~al.}(2015{\natexlab{b}})\citenamefont {Chen}, \citenamefont {Chen},
  \citenamefont {Liu}, \citenamefont {Steele},\ and\ \citenamefont
  {Zhu}}]{Chen:2015moa}%
  \BibitemOpen
  \bibfield  {author} {\bibinfo {author} {\bibfnamefont {H.-X.}\ \bibnamefont
  {Chen}}, \bibinfo {author} {\bibfnamefont {W.}~\bibnamefont {Chen}}, \bibinfo
  {author} {\bibfnamefont {X.}~\bibnamefont {Liu}}, \bibinfo {author}
  {\bibfnamefont {T.~G.}\ \bibnamefont {Steele}}, \ and\ \bibinfo {author}
  {\bibfnamefont {S.-L.}\ \bibnamefont {Zhu}},\ }\href {\doibase
  10.1103/PhysRevLett.115.172001} {\bibfield  {journal} {\bibinfo  {journal}
  {Phys. Rev. Lett.}\ }\textbf {\bibinfo {volume} {115}},\ \bibinfo {pages}
  {172001} (\bibinfo {year} {2015}{\natexlab{b}})},\ \Eprint
  {http://arxiv.org/abs/1507.03717} {arXiv:1507.03717 [hep-ph]} \BibitemShut
  {NoStop}%
%%CITATION = ARXIV:1507.03717;%%
\bibitem [{\citenamefont {Karliner}\ and\ \citenamefont
  {Rosner}(2015)}]{Karliner:2015ina}%
  \BibitemOpen
  \bibfield  {author} {\bibinfo {author} {\bibfnamefont {M.}~\bibnamefont
  {Karliner}}\ and\ \bibinfo {author} {\bibfnamefont {J.~L.}\ \bibnamefont
  {Rosner}},\ }\href {\doibase 10.1103/PhysRevLett.115.122001} {\bibfield
  {journal} {\bibinfo  {journal} {Phys. Rev. Lett.}\ }\textbf {\bibinfo
  {volume} {115}},\ \bibinfo {pages} {122001} (\bibinfo {year} {2015})},\
  \Eprint {http://arxiv.org/abs/1506.06386} {arXiv:1506.06386 [hep-ph]}
  \BibitemShut {NoStop}%
%%CITATION = ARXIV:1506.06386;%%
\bibitem [{\citenamefont {Roca}\ \emph {et~al.}(2015)\citenamefont {Roca},
  \citenamefont {Nieves},\ and\ \citenamefont {Oset}}]{Roca:2015dva}%
  \BibitemOpen
  \bibfield  {author} {\bibinfo {author} {\bibfnamefont {L.}~\bibnamefont
  {Roca}}, \bibinfo {author} {\bibfnamefont {J.}~\bibnamefont {Nieves}}, \ and\
  \bibinfo {author} {\bibfnamefont {E.}~\bibnamefont {Oset}},\ }\href {\doibase
  10.1103/PhysRevD.92.094003} {\bibfield  {journal} {\bibinfo  {journal} {Phys.
  Rev.}\ }\textbf {\bibinfo {volume} {D92}},\ \bibinfo {pages} {094003}
  (\bibinfo {year} {2015})},\ \Eprint {http://arxiv.org/abs/1507.04249}
  {arXiv:1507.04249 [hep-ph]} \BibitemShut {NoStop}%
%%CITATION = ARXIV:1507.04249;%%
\bibitem [{\citenamefont {He}(2016)}]{He:2015cea}%
  \BibitemOpen
  \bibfield  {author} {\bibinfo {author} {\bibfnamefont {J.}~\bibnamefont
  {He}},\ }\href {\doibase 10.1016/j.physletb.2015.12.071} {\bibfield
  {journal} {\bibinfo  {journal} {Phys. Lett.}\ }\textbf {\bibinfo {volume}
  {B753}},\ \bibinfo {pages} {547} (\bibinfo {year} {2016})},\ \Eprint
  {http://arxiv.org/abs/1507.05200} {arXiv:1507.05200 [hep-ph]} \BibitemShut
  {NoStop}%
%%CITATION = ARXIV:1507.05200;%%
\bibitem [{\citenamefont {Mei?ner}\ and\ \citenamefont
  {Oller}(2015)}]{Meissner:2015mza}%
  \BibitemOpen
  \bibfield  {author} {\bibinfo {author} {\bibfnamefont {U.-G.}\ \bibnamefont
  {Mei?ner}}\ and\ \bibinfo {author} {\bibfnamefont {J.~A.}\ \bibnamefont
  {Oller}},\ }\href {\doibase 10.1016/j.physletb.2015.10.015} {\bibfield
  {journal} {\bibinfo  {journal} {Phys. Lett.}\ }\textbf {\bibinfo {volume}
  {B751}},\ \bibinfo {pages} {59} (\bibinfo {year} {2015})},\ \Eprint
  {http://arxiv.org/abs/1507.07478} {arXiv:1507.07478 [hep-ph]} \BibitemShut
  {NoStop}%
%%CITATION = ARXIV:1507.07478;%%
\bibitem [{\citenamefont {Shen}\ \emph {et~al.}(2016)\citenamefont {Shen},
  \citenamefont {Guo}, \citenamefont {Xie},\ and\ \citenamefont
  {Zou}}]{Shen:2016tzq}%
  \BibitemOpen
  \bibfield  {author} {\bibinfo {author} {\bibfnamefont {C.-W.}\ \bibnamefont
  {Shen}}, \bibinfo {author} {\bibfnamefont {F.-K.}\ \bibnamefont {Guo}},
  \bibinfo {author} {\bibfnamefont {J.-J.}\ \bibnamefont {Xie}}, \ and\
  \bibinfo {author} {\bibfnamefont {B.-S.}\ \bibnamefont {Zou}},\ }\href
  {\doibase 10.1016/j.nuclphysa.2016.04.034} {\bibfield  {journal} {\bibinfo
  {journal} {Nucl. Phys.}\ }\textbf {\bibinfo {volume} {A954}},\ \bibinfo
  {pages} {393} (\bibinfo {year} {2016})},\ \Eprint
  {http://arxiv.org/abs/1603.04672} {arXiv:1603.04672 [hep-ph]} \BibitemShut
  {NoStop}%
%%CITATION = ARXIV:1603.04672;%%
\bibitem [{\citenamefont {Liu}\ \emph {et~al.}(2019{\natexlab{a}})\citenamefont
  {Liu}, \citenamefont {Pan}, \citenamefont {Peng}, \citenamefont
  {S¨¢nchez~S¨¢nchez}, \citenamefont {Geng}, \citenamefont {Hosaka},\ and\
  \citenamefont {Pavon~Valderrama}}]{Liu:2019tjn}%
  \BibitemOpen
  \bibfield  {author} {\bibinfo {author} {\bibfnamefont {M.-Z.}\ \bibnamefont
  {Liu}}, \bibinfo {author} {\bibfnamefont {Y.-W.}\ \bibnamefont {Pan}},
  \bibinfo {author} {\bibfnamefont {F.-Z.}\ \bibnamefont {Peng}}, \bibinfo
  {author} {\bibfnamefont {M.}~\bibnamefont {S¨¢nchez~S¨¢nchez}}, \bibinfo
  {author} {\bibfnamefont {L.-S.}\ \bibnamefont {Geng}}, \bibinfo {author}
  {\bibfnamefont {A.}~\bibnamefont {Hosaka}}, \ and\ \bibinfo {author}
  {\bibfnamefont {M.}~\bibnamefont {Pavon~Valderrama}},\ }\href {\doibase
  10.1103/PhysRevLett.122.242001} {\bibfield  {journal} {\bibinfo  {journal}
  {Phys. Rev. Lett.}\ }\textbf {\bibinfo {volume} {122}},\ \bibinfo {pages}
  {242001} (\bibinfo {year} {2019}{\natexlab{a}})},\ \Eprint
  {http://arxiv.org/abs/1903.11560} {arXiv:1903.11560 [hep-ph]} \BibitemShut
  {NoStop}%
%%CITATION = ARXIV:1903.11560;%%
\bibitem [{\citenamefont {Maiani}\ \emph {et~al.}(2015)\citenamefont {Maiani},
  \citenamefont {Polosa},\ and\ \citenamefont {Riquer}}]{Maiani:2015vwa}%
  \BibitemOpen
  \bibfield  {author} {\bibinfo {author} {\bibfnamefont {L.}~\bibnamefont
  {Maiani}}, \bibinfo {author} {\bibfnamefont {A.~D.}\ \bibnamefont {Polosa}},
  \ and\ \bibinfo {author} {\bibfnamefont {V.}~\bibnamefont {Riquer}},\ }\href
  {\doibase 10.1016/j.physletb.2015.08.008} {\bibfield  {journal} {\bibinfo
  {journal} {Phys. Lett.}\ }\textbf {\bibinfo {volume} {B749}},\ \bibinfo
  {pages} {289} (\bibinfo {year} {2015})},\ \Eprint
  {http://arxiv.org/abs/1507.04980} {arXiv:1507.04980 [hep-ph]} \BibitemShut
  {NoStop}%
%%CITATION = ARXIV:1507.04980;%%
\bibitem [{\citenamefont {Wang}\ and\ \citenamefont
  {Huang}(2016)}]{Wang:2015ava}%
  \BibitemOpen
  \bibfield  {author} {\bibinfo {author} {\bibfnamefont {Z.-G.}\ \bibnamefont
  {Wang}}\ and\ \bibinfo {author} {\bibfnamefont {T.}~\bibnamefont {Huang}},\
  }\href {\doibase 10.1140/epjc/s10052-016-3880-8} {\bibfield  {journal}
  {\bibinfo  {journal} {Eur. Phys. J.}\ }\textbf {\bibinfo {volume} {C76}},\
  \bibinfo {pages} {43} (\bibinfo {year} {2016})},\ \Eprint
  {http://arxiv.org/abs/1508.04189} {arXiv:1508.04189 [hep-ph]} \BibitemShut
  {NoStop}%
%%CITATION = ARXIV:1508.04189;%%
\bibitem [{\citenamefont {Lebed}(2015)}]{Lebed:2015tna}%
  \BibitemOpen
  \bibfield  {author} {\bibinfo {author} {\bibfnamefont {R.~F.}\ \bibnamefont
  {Lebed}},\ }\href {\doibase 10.1016/j.physletb.2015.08.032} {\bibfield
  {journal} {\bibinfo  {journal} {Phys. Lett.}\ }\textbf {\bibinfo {volume}
  {B749}},\ \bibinfo {pages} {454} (\bibinfo {year} {2015})},\ \Eprint
  {http://arxiv.org/abs/1507.05867} {arXiv:1507.05867 [hep-ph]} \BibitemShut
  {NoStop}%
%%CITATION = ARXIV:1507.05867;%%
\bibitem [{\citenamefont {Wang}(2016)}]{Wang:2015epa}%
  \BibitemOpen
  \bibfield  {author} {\bibinfo {author} {\bibfnamefont {Z.-G.}\ \bibnamefont
  {Wang}},\ }\href {\doibase 10.1140/epjc/s10052-016-3920-4} {\bibfield
  {journal} {\bibinfo  {journal} {Eur. Phys. J.}\ }\textbf {\bibinfo {volume}
  {C76}},\ \bibinfo {pages} {70} (\bibinfo {year} {2016})},\ \Eprint
  {http://arxiv.org/abs/1508.01468} {arXiv:1508.01468 [hep-ph]} \BibitemShut
  {NoStop}%
%%CITATION = ARXIV:1508.01468;%%
\bibitem [{\citenamefont {Guo}\ \emph {et~al.}(2015)\citenamefont {Guo},
  \citenamefont {Mei?ner}, \citenamefont {Wang},\ and\ \citenamefont
  {Yang}}]{Guo:2015umn}%
  \BibitemOpen
  \bibfield  {author} {\bibinfo {author} {\bibfnamefont {F.-K.}\ \bibnamefont
  {Guo}}, \bibinfo {author} {\bibfnamefont {U.-G.}\ \bibnamefont {Mei?ner}},
  \bibinfo {author} {\bibfnamefont {W.}~\bibnamefont {Wang}}, \ and\ \bibinfo
  {author} {\bibfnamefont {Z.}~\bibnamefont {Yang}},\ }\href {\doibase
  10.1103/PhysRevD.92.071502} {\bibfield  {journal} {\bibinfo  {journal} {Phys.
  Rev.}\ }\textbf {\bibinfo {volume} {D92}},\ \bibinfo {pages} {071502}
  (\bibinfo {year} {2015})},\ \Eprint {http://arxiv.org/abs/1507.04950}
  {arXiv:1507.04950 [hep-ph]} \BibitemShut {NoStop}%
%%CITATION = ARXIV:1507.04950;%%
\bibitem [{\citenamefont {Liu}\ \emph {et~al.}(2016)\citenamefont {Liu},
  \citenamefont {Wang},\ and\ \citenamefont {Zhao}}]{Liu:2015fea}%
  \BibitemOpen
  \bibfield  {author} {\bibinfo {author} {\bibfnamefont {X.-H.}\ \bibnamefont
  {Liu}}, \bibinfo {author} {\bibfnamefont {Q.}~\bibnamefont {Wang}}, \ and\
  \bibinfo {author} {\bibfnamefont {Q.}~\bibnamefont {Zhao}},\ }\href {\doibase
  10.1016/j.physletb.2016.03.089} {\bibfield  {journal} {\bibinfo  {journal}
  {Phys. Lett.}\ }\textbf {\bibinfo {volume} {B757}},\ \bibinfo {pages} {231}
  (\bibinfo {year} {2016})},\ \Eprint {http://arxiv.org/abs/1507.05359}
  {arXiv:1507.05359 [hep-ph]} \BibitemShut {NoStop}%
%%CITATION = ARXIV:1507.05359;%%
\bibitem [{\citenamefont {Scoccola}\ \emph {et~al.}(2015)\citenamefont
  {Scoccola}, \citenamefont {Riska},\ and\ \citenamefont
  {Rho}}]{Scoccola:2015nia}%
  \BibitemOpen
  \bibfield  {author} {\bibinfo {author} {\bibfnamefont {N.~N.}\ \bibnamefont
  {Scoccola}}, \bibinfo {author} {\bibfnamefont {D.~O.}\ \bibnamefont {Riska}},
  \ and\ \bibinfo {author} {\bibfnamefont {M.}~\bibnamefont {Rho}},\ }\href
  {\doibase 10.1103/PhysRevD.92.051501} {\bibfield  {journal} {\bibinfo
  {journal} {Phys. Rev.}\ }\textbf {\bibinfo {volume} {D92}},\ \bibinfo {pages}
  {051501} (\bibinfo {year} {2015})},\ \Eprint
  {http://arxiv.org/abs/1508.01172} {arXiv:1508.01172 [hep-ph]} \BibitemShut
  {NoStop}%
%%CITATION = ARXIV:1508.01172;%%
\bibitem [{\citenamefont {Chen}\ \emph {et~al.}(2016)\citenamefont {Chen},
  \citenamefont {Chen}, \citenamefont {Liu},\ and\ \citenamefont
  {Zhu}}]{Chen:2016qju}%
  \BibitemOpen
  \bibfield  {author} {\bibinfo {author} {\bibfnamefont {H.-X.}\ \bibnamefont
  {Chen}}, \bibinfo {author} {\bibfnamefont {W.}~\bibnamefont {Chen}}, \bibinfo
  {author} {\bibfnamefont {X.}~\bibnamefont {Liu}}, \ and\ \bibinfo {author}
  {\bibfnamefont {S.-L.}\ \bibnamefont {Zhu}},\ }\href {\doibase
  10.1016/j.physrep.2016.05.004} {\bibfield  {journal} {\bibinfo  {journal}
  {Phys. Rept.}\ }\textbf {\bibinfo {volume} {639}},\ \bibinfo {pages} {1}
  (\bibinfo {year} {2016})},\ \Eprint {http://arxiv.org/abs/1601.02092}
  {arXiv:1601.02092 [hep-ph]} \BibitemShut {NoStop}%
%%CITATION = ARXIV:1601.02092;%%
\bibitem [{\citenamefont {Ali}\ \emph {et~al.}(2017)\citenamefont {Ali},
  \citenamefont {Lange},\ and\ \citenamefont {Stone}}]{Ali:2017jda}%
  \BibitemOpen
  \bibfield  {author} {\bibinfo {author} {\bibfnamefont {A.}~\bibnamefont
  {Ali}}, \bibinfo {author} {\bibfnamefont {J.~S.}\ \bibnamefont {Lange}}, \
  and\ \bibinfo {author} {\bibfnamefont {S.}~\bibnamefont {Stone}},\ }\href
  {\doibase 10.1016/j.ppnp.2017.08.003} {\bibfield  {journal} {\bibinfo
  {journal} {Prog. Part. Nucl. Phys.}\ }\textbf {\bibinfo {volume} {97}},\
  \bibinfo {pages} {123} (\bibinfo {year} {2017})},\ \Eprint
  {http://arxiv.org/abs/1706.00610} {arXiv:1706.00610 [hep-ph]} \BibitemShut
  {NoStop}%
%%CITATION = ARXIV:1706.00610;%%
\bibitem [{\citenamefont {Guo}\ \emph {et~al.}(2018)\citenamefont {Guo},
  \citenamefont {Hanhart}, \citenamefont {Mei?ner}, \citenamefont {Wang},
  \citenamefont {Zhao},\ and\ \citenamefont {Zou}}]{Guo:2017jvc}%
  \BibitemOpen
  \bibfield  {author} {\bibinfo {author} {\bibfnamefont {F.-K.}\ \bibnamefont
  {Guo}}, \bibinfo {author} {\bibfnamefont {C.}~\bibnamefont {Hanhart}},
  \bibinfo {author} {\bibfnamefont {U.-G.}\ \bibnamefont {Mei?ner}}, \bibinfo
  {author} {\bibfnamefont {Q.}~\bibnamefont {Wang}}, \bibinfo {author}
  {\bibfnamefont {Q.}~\bibnamefont {Zhao}}, \ and\ \bibinfo {author}
  {\bibfnamefont {B.-S.}\ \bibnamefont {Zou}},\ }\href {\doibase
  10.1103/RevModPhys.90.015004} {\bibfield  {journal} {\bibinfo  {journal}
  {Rev. Mod. Phys.}\ }\textbf {\bibinfo {volume} {90}},\ \bibinfo {pages}
  {015004} (\bibinfo {year} {2018})},\ \Eprint
  {http://arxiv.org/abs/1705.00141} {arXiv:1705.00141 [hep-ph]} \BibitemShut
  {NoStop}%
%%CITATION = ARXIV:1705.00141;%%
\bibitem [{\citenamefont {Liu}\ \emph {et~al.}(2019{\natexlab{b}})\citenamefont
  {Liu}, \citenamefont {Chen}, \citenamefont {Chen}, \citenamefont {Liu},\ and\
  \citenamefont {Zhu}}]{Liu:2019zoy}%
  \BibitemOpen
  \bibfield  {author} {\bibinfo {author} {\bibfnamefont {Y.-R.}\ \bibnamefont
  {Liu}}, \bibinfo {author} {\bibfnamefont {H.-X.}\ \bibnamefont {Chen}},
  \bibinfo {author} {\bibfnamefont {W.}~\bibnamefont {Chen}}, \bibinfo {author}
  {\bibfnamefont {X.}~\bibnamefont {Liu}}, \ and\ \bibinfo {author}
  {\bibfnamefont {S.-L.}\ \bibnamefont {Zhu}},\ }\href {\doibase
  10.1016/j.ppnp.2019.04.003} {\bibfield  {journal} {\bibinfo  {journal} {Prog.
  Part. Nucl. Phys.}\ }\textbf {\bibinfo {volume} {107}},\ \bibinfo {pages}
  {237} (\bibinfo {year} {2019}{\natexlab{b}})},\ \Eprint
  {http://arxiv.org/abs/1903.11976} {arXiv:1903.11976 [hep-ph]} \BibitemShut
  {NoStop}%
%%CITATION = ARXIV:1903.11976;%%
\bibitem [{\citenamefont {Nowak}\ \emph {et~al.}(1993)\citenamefont {Nowak},
  \citenamefont {Rho},\ and\ \citenamefont {Zahed}}]{Nowak:1992um}%
  \BibitemOpen
  \bibfield  {author} {\bibinfo {author} {\bibfnamefont {M.~A.}\ \bibnamefont
  {Nowak}}, \bibinfo {author} {\bibfnamefont {M.}~\bibnamefont {Rho}}, \ and\
  \bibinfo {author} {\bibfnamefont {I.}~\bibnamefont {Zahed}},\ }\href
  {\doibase 10.1103/PhysRevD.48.4370} {\bibfield  {journal} {\bibinfo
  {journal} {Phys. Rev.}\ }\textbf {\bibinfo {volume} {D48}},\ \bibinfo {pages}
  {4370} (\bibinfo {year} {1993})},\ \Eprint
  {http://arxiv.org/abs/hep-ph/9209272} {arXiv:hep-ph/9209272 [hep-ph]}
  \BibitemShut {NoStop}%
%%CITATION = HEP-PH/9209272;%%
\bibitem [{\citenamefont {Bardeen}\ and\ \citenamefont
  {Hill}(1994)}]{Bardeen:1993ae}%
  \BibitemOpen
  \bibfield  {author} {\bibinfo {author} {\bibfnamefont {W.~A.}\ \bibnamefont
  {Bardeen}}\ and\ \bibinfo {author} {\bibfnamefont {C.~T.}\ \bibnamefont
  {Hill}},\ }\href {\doibase 10.1103/PhysRevD.49.409} {\bibfield  {journal}
  {\bibinfo  {journal} {Phys. Rev.}\ }\textbf {\bibinfo {volume} {D49}},\
  \bibinfo {pages} {409} (\bibinfo {year} {1994})},\ \Eprint
  {http://arxiv.org/abs/hep-ph/9304265} {arXiv:hep-ph/9304265 [hep-ph]}
  \BibitemShut {NoStop}%
%%CITATION = HEP-PH/9304265;%%
\bibitem [{\citenamefont {Harada}\ and\ \citenamefont
  {Ma}(2013)}]{Harada:2012dm}%
  \BibitemOpen
  \bibfield  {author} {\bibinfo {author} {\bibfnamefont {M.}~\bibnamefont
  {Harada}}\ and\ \bibinfo {author} {\bibfnamefont {Y.-L.}\ \bibnamefont
  {Ma}},\ }\href {\doibase 10.1103/PhysRevD.87.056007} {\bibfield  {journal}
  {\bibinfo  {journal} {Phys. Rev.}\ }\textbf {\bibinfo {volume} {D87}},\
  \bibinfo {pages} {056007} (\bibinfo {year} {2013})},\ \Eprint
  {http://arxiv.org/abs/1212.5079} {arXiv:1212.5079 [hep-ph]} \BibitemShut
  {NoStop}%
%%CITATION = ARXIV:1212.5079;%%
\bibitem [{\citenamefont {Oh}\ \emph {et~al.}(1994{\natexlab{a}})\citenamefont
  {Oh}, \citenamefont {Park},\ and\ \citenamefont {Min}}]{Oh:1994yv}%
  \BibitemOpen
  \bibfield  {author} {\bibinfo {author} {\bibfnamefont {Y.-s.}\ \bibnamefont
  {Oh}}, \bibinfo {author} {\bibfnamefont {B.-Y.}\ \bibnamefont {Park}}, \ and\
  \bibinfo {author} {\bibfnamefont {D.-P.}\ \bibnamefont {Min}},\ }\href
  {\doibase 10.1103/PhysRevD.50.3350} {\bibfield  {journal} {\bibinfo
  {journal} {Phys. Rev.}\ }\textbf {\bibinfo {volume} {D50}},\ \bibinfo {pages}
  {3350} (\bibinfo {year} {1994}{\natexlab{a}})},\ \Eprint
  {http://arxiv.org/abs/hep-ph/9407214} {arXiv:hep-ph/9407214 [hep-ph]}
  \BibitemShut {NoStop}%
%%CITATION = HEP-PH/9407214;%%
\bibitem [{\citenamefont {Oh}\ \emph {et~al.}(1994{\natexlab{b}})\citenamefont
  {Oh}, \citenamefont {Park},\ and\ \citenamefont {Min}}]{Oh:1994np}%
  \BibitemOpen
  \bibfield  {author} {\bibinfo {author} {\bibfnamefont {Y.-s.}\ \bibnamefont
  {Oh}}, \bibinfo {author} {\bibfnamefont {B.-Y.}\ \bibnamefont {Park}}, \ and\
  \bibinfo {author} {\bibfnamefont {D.-P.}\ \bibnamefont {Min}},\ }\href
  {\doibase 10.1016/0370-2693(94)91065-0} {\bibfield  {journal} {\bibinfo
  {journal} {Phys. Lett.}\ }\textbf {\bibinfo {volume} {B331}},\ \bibinfo
  {pages} {362} (\bibinfo {year} {1994}{\natexlab{b}})},\ \Eprint
  {http://arxiv.org/abs/hep-ph/9405297} {arXiv:hep-ph/9405297 [hep-ph]}
  \BibitemShut {NoStop}%
%%CITATION = HEP-PH/9405297;%%
\bibitem [{\citenamefont {Harada}\ \emph {et~al.}(1997)\citenamefont {Harada},
  \citenamefont {Sannino}, \citenamefont {Schechter},\ and\ \citenamefont
  {Weigel}}]{Harada:1997bk}%
  \BibitemOpen
  \bibfield  {author} {\bibinfo {author} {\bibfnamefont {M.}~\bibnamefont
  {Harada}}, \bibinfo {author} {\bibfnamefont {F.}~\bibnamefont {Sannino}},
  \bibinfo {author} {\bibfnamefont {J.}~\bibnamefont {Schechter}}, \ and\
  \bibinfo {author} {\bibfnamefont {H.}~\bibnamefont {Weigel}},\ }\href
  {\doibase 10.1103/PhysRevD.56.4098} {\bibfield  {journal} {\bibinfo
  {journal} {Phys. Rev.}\ }\textbf {\bibinfo {volume} {D56}},\ \bibinfo {pages}
  {4098} (\bibinfo {year} {1997})},\ \Eprint
  {http://arxiv.org/abs/hep-ph/9704358} {arXiv:hep-ph/9704358 [hep-ph]}
  \BibitemShut {NoStop}%
%%CITATION = HEP-PH/9704358;%%
\bibitem [{\citenamefont {Wu}\ and\ \citenamefont {Ma}(2004)}]{Wu:2004wg}%
  \BibitemOpen
  \bibfield  {author} {\bibinfo {author} {\bibfnamefont {B.}~\bibnamefont
  {Wu}}\ and\ \bibinfo {author} {\bibfnamefont {B.-Q.}\ \bibnamefont {Ma}},\
  }\href {\doibase 10.1103/PhysRevD.70.034025} {\bibfield  {journal} {\bibinfo
  {journal} {Phys. Rev.}\ }\textbf {\bibinfo {volume} {D70}},\ \bibinfo {pages}
  {034025} (\bibinfo {year} {2004})},\ \Eprint
  {http://arxiv.org/abs/hep-ph/0402244} {arXiv:hep-ph/0402244 [hep-ph]}
  \BibitemShut {NoStop}%
%%CITATION = HEP-PH/0402244;%%
\bibitem [{\citenamefont {Schechter}\ \emph {et~al.}(1995)\citenamefont
  {Schechter}, \citenamefont {Subbaraman}, \citenamefont {Vaidya},\ and\
  \citenamefont {Weigel}}]{Schechter:1995vr}%
  \BibitemOpen
  \bibfield  {author} {\bibinfo {author} {\bibfnamefont {J.}~\bibnamefont
  {Schechter}}, \bibinfo {author} {\bibfnamefont {A.}~\bibnamefont
  {Subbaraman}}, \bibinfo {author} {\bibfnamefont {S.}~\bibnamefont {Vaidya}},
  \ and\ \bibinfo {author} {\bibfnamefont {H.}~\bibnamefont {Weigel}},\ }\href
  {\doibase 10.1016/0375-9474(95)00182-Z, 10.1016/0375-9474(96)00013-9}
  {\bibfield  {journal} {\bibinfo  {journal} {Nucl. Phys.}\ }\textbf {\bibinfo
  {volume} {A590}},\ \bibinfo {pages} {655} (\bibinfo {year} {1995})},\
  \bibinfo {note} {[Erratum: Nucl. Phys.A598,583(1996)]},\ \Eprint
  {http://arxiv.org/abs/hep-ph/9503307} {arXiv:hep-ph/9503307 [hep-ph]}
  \BibitemShut {NoStop}%
%%CITATION = HEP-PH/9503307;%%
\bibitem [{\citenamefont {Nowak}\ \emph {et~al.}(2004)\citenamefont {Nowak},
  \citenamefont {Praszalowicz}, \citenamefont {Sadzikowski},\ and\
  \citenamefont {Wasiluk}}]{Nowak:2004jg}%
  \BibitemOpen
  \bibfield  {author} {\bibinfo {author} {\bibfnamefont {M.~A.}\ \bibnamefont
  {Nowak}}, \bibinfo {author} {\bibfnamefont {M.}~\bibnamefont {Praszalowicz}},
  \bibinfo {author} {\bibfnamefont {M.}~\bibnamefont {Sadzikowski}}, \ and\
  \bibinfo {author} {\bibfnamefont {J.}~\bibnamefont {Wasiluk}},\ }\href
  {\doibase 10.1103/PhysRevD.70.031503} {\bibfield  {journal} {\bibinfo
  {journal} {Phys. Rev.}\ }\textbf {\bibinfo {volume} {D70}},\ \bibinfo {pages}
  {031503} (\bibinfo {year} {2004})},\ \Eprint
  {http://arxiv.org/abs/hep-ph/0403184} {arXiv:hep-ph/0403184 [hep-ph]}
  \BibitemShut {NoStop}%
%%CITATION = HEP-PH/0403184;%%
\bibitem [{\citenamefont {Harada}\ and\ \citenamefont
  {Yamawaki}(2003)}]{Harada:2003jx}%
  \BibitemOpen
  \bibfield  {author} {\bibinfo {author} {\bibfnamefont {M.}~\bibnamefont
  {Harada}}\ and\ \bibinfo {author} {\bibfnamefont {K.}~\bibnamefont
  {Yamawaki}},\ }\href {\doibase 10.1016/S0370-1573(03)00139-X} {\bibfield
  {journal} {\bibinfo  {journal} {Phys. Rept.}\ }\textbf {\bibinfo {volume}
  {381}},\ \bibinfo {pages} {1} (\bibinfo {year} {2003})},\ \Eprint
  {http://arxiv.org/abs/hep-ph/0302103} {arXiv:hep-ph/0302103 [hep-ph]}
  \BibitemShut {NoStop}%
%%CITATION = HEP-PH/0302103;%%
\bibitem [{\citenamefont {Tanabashi}\ \emph {et~al.}(2018)\citenamefont
  {Tanabashi} \emph {et~al.}}]{Tanabashi:2018oca}%
  \BibitemOpen
  \bibfield  {author} {\bibinfo {author} {\bibfnamefont {M.}~\bibnamefont
  {Tanabashi}} \emph {et~al.} (\bibinfo {collaboration} {Particle Data
  Group}),\ }\href {\doibase 10.1103/PhysRevD.98.030001} {\bibfield  {journal}
  {\bibinfo  {journal} {Phys. Rev.}\ }\textbf {\bibinfo {volume} {D98}},\
  \bibinfo {pages} {030001} (\bibinfo {year} {2018})}\BibitemShut {NoStop}%
%%CITATION = PHRVA,D98,030001;%%
\bibitem [{\citenamefont {Skyrme}(1961)}]{Skyrme:1961vq}%
  \BibitemOpen
  \bibfield  {author} {\bibinfo {author} {\bibfnamefont {T.~H.~R.}\
  \bibnamefont {Skyrme}},\ }\href {\doibase 10.1098/rspa.1961.0018} {\bibfield
  {journal} {\bibinfo  {journal} {Proc. Roy. Soc. Lond.}\ }\textbf {\bibinfo
  {volume} {A260}},\ \bibinfo {pages} {127} (\bibinfo {year}
  {1961})}\BibitemShut {NoStop}%
%%CITATION = PRSLA,A260,127;%%
\bibitem [{\citenamefont {Ma}\ \emph {et~al.}(2013)\citenamefont {Ma},
  \citenamefont {Yang}, \citenamefont {Oh},\ and\ \citenamefont
  {Harada}}]{Ma:2012zm}%
  \BibitemOpen
  \bibfield  {author} {\bibinfo {author} {\bibfnamefont {Y.-L.}\ \bibnamefont
  {Ma}}, \bibinfo {author} {\bibfnamefont {G.-S.}\ \bibnamefont {Yang}},
  \bibinfo {author} {\bibfnamefont {Y.}~\bibnamefont {Oh}}, \ and\ \bibinfo
  {author} {\bibfnamefont {M.}~\bibnamefont {Harada}},\ }\href {\doibase
  10.1103/PhysRevD.87.034023} {\bibfield  {journal} {\bibinfo  {journal} {Phys.
  Rev.}\ }\textbf {\bibinfo {volume} {D87}},\ \bibinfo {pages} {034023}
  (\bibinfo {year} {2013})},\ \Eprint {http://arxiv.org/abs/1209.3554}
  {arXiv:1209.3554 [hep-ph]} \BibitemShut {NoStop}%
%%CITATION = ARXIV:1209.3554;%%
\bibitem [{\citenamefont {Callan}\ and\ \citenamefont
  {Klebanov}(1985)}]{Callan:1985hy}%
  \BibitemOpen
  \bibfield  {author} {\bibinfo {author} {\bibfnamefont {C.~G.}\ \bibnamefont
  {Callan}, \bibfnamefont {Jr.}}\ and\ \bibinfo {author} {\bibfnamefont
  {I.~R.}\ \bibnamefont {Klebanov}},\ }\href {\doibase
  10.1016/0550-3213(85)90292-5} {\bibfield  {journal} {\bibinfo  {journal}
  {Nucl. Phys.}\ }\textbf {\bibinfo {volume} {B262}},\ \bibinfo {pages} {365}
  (\bibinfo {year} {1985})}\BibitemShut {NoStop}%
%%CITATION = NUPHA,B262,365;%%
\bibitem [{\citenamefont {Harada}\ \emph {et~al.}(2012)\citenamefont {Harada},
  \citenamefont {Hoshino},\ and\ \citenamefont {Ma}}]{Harada:2012km}%
  \BibitemOpen
  \bibfield  {author} {\bibinfo {author} {\bibfnamefont {M.}~\bibnamefont
  {Harada}}, \bibinfo {author} {\bibfnamefont {H.}~\bibnamefont {Hoshino}}, \
  and\ \bibinfo {author} {\bibfnamefont {Y.-L.}\ \bibnamefont {Ma}},\ }\href
  {\doibase 10.1103/PhysRevD.85.114027} {\bibfield  {journal} {\bibinfo
  {journal} {Phys. Rev.}\ }\textbf {\bibinfo {volume} {D85}},\ \bibinfo {pages}
  {114027} (\bibinfo {year} {2012})},\ \Eprint {http://arxiv.org/abs/1203.3632}
  {arXiv:1203.3632 [hep-ph]} \BibitemShut {NoStop}%
%%CITATION = ARXIV:1203.3632;%%
\bibitem [{\citenamefont {Rho}\ \emph {et~al.}(1990)\citenamefont {Rho},
  \citenamefont {Riska},\ and\ \citenamefont {Scoccola}}]{Rho:1990uy}%
  \BibitemOpen
  \bibfield  {author} {\bibinfo {author} {\bibfnamefont {M.}~\bibnamefont
  {Rho}}, \bibinfo {author} {\bibfnamefont {D.~O.}\ \bibnamefont {Riska}}, \
  and\ \bibinfo {author} {\bibfnamefont {N.~N.}\ \bibnamefont {Scoccola}},\
  }\href {\doibase 10.1016/0370-2693(90)90802-D} {\bibfield  {journal}
  {\bibinfo  {journal} {Phys. Lett.}\ }\textbf {\bibinfo {volume} {B251}},\
  \bibinfo {pages} {597} (\bibinfo {year} {1990})}\BibitemShut {NoStop}%
%%CITATION = PHLTA,B251,597;%%
\end{thebibliography}%

\end{document}